\documentstyle[art12]{article}
\begin{document}

\title {MAGNETOSPHERE-IONOSPHERE COUPLING \\
IN THE REGION OF AURORAL ELECTROJETS}

\author { P.A.Sedykh and E.A.Ponomarev \\
Institute of Solar-Terrestrial Physics SD RAS, Irkutsk, Russia}

\maketitle

Abstract. We discuss the question as to how the magnetospheric energy source feeds 
the ionospheric current system. It is shown that a consistent application and further 
development of Kennel's ideas makes it possible to successfully solve the 
magnetosphere-ionosphere coupling problem in regard to the formation of auroral 
electrojets by steady volume currents generated in the magnetosphere by the 
magnetospheric MHD generator in the case of a simple model which, nevertheless, 
retains the essential features of the reality. It is concluded that the whole of the 
complicated magnetospheric ''design'' only acts to redistribute, in space and time, 
currents and energy fluxes which must be supplied by external sources to feed the 
dissipative processes in the ionosphere.
\\

\section{INTRODUCTION}

The resumes of the last two international conferences [Kamide, 1998; Lui, 2000] 
distinctly voiced a faint note of dissatisfaction with the lack of progress toward an 
understanding of the physical essence of magnetospheric processes and, above all, the 
magnetospheric substorm. In our opinion, to speak in plain terms, things are reaching a 
conceptual crisis. Four decades ago, two concepts of solar wind (SW)-magnetosphere 
coupling were formulated: the concept of quasi-viscous interaction [Axford and Hines, 
1961], and the concept of magnetic field line reconnection [Dungey, 1961]. The 
concepts are both based on the assumption that in the region of SW-magnetosphere 
interaction there exist meaningful transport coefficients: a certain effective viscosity 
and effective conductivity in the former and latter cases, respectively. These 
coefficients are both proportional to a certain length having the meaning of the free path 
length of particles in the case of paired collisions. Since the free path length of 
particles in the magnetosphere at paired collisions involving Coulomb interaction is far 
in excess of the dimensions of the magnetosphere, magnetospheric plasma is said to be 
collisionless. However, plasma can sustain collective processes leading to exchanges 
of momentum and energy between particles. The exchange proceeds through waves 
which must have a spectral energy density large enough for a sufficient exchange rate to 
be ensured (quasi-collisional mode). Thus the validity of both concepts boils down to 
the problem of finding out such plasma instabilities which would be capable of 
ensuring the quasi-collisional regime. This problem has not been solved yet, and hence 
groundwork is lacking for both concepts of energy transfer from the solar wind to the 
magnetosphere. Still this does not seem to be the main reason behind the crisis. The 
required instabilities may well be found. Moreover, recently V.V.Mishin [Mishin, 
2001] was able to demonstrate that when supersonic magnetosheeth plasma flows 
round the magnetopause, the shear flow generates oblique magnetosonic waves 
penetrating deep into the magnetosphere and carrying their momentum along. This 
process proved to be sufficiently effective, so it must be taken into account in the 
energy budget of the magnetosphere. The chief reason, however, is that neither of the 
two concepts failed to serve as an appropriate foundation for constructing a sequence 
of physical mechanisms which would lead us from the processes in the bow shock to 
the processes of auroral electrojet formation.\\
     To find a way out of the impasse implies  adopting a new concept based on the 
well-known Kennel's paper entitled ''Consequences of magnetospheric plasma'' 
[Kennel, 1969]. The essentials of this concept may be summarized as follows. The 
combined action of convection and pitch-angle diffusion leads to the formation in the 
magnetosphere of a spatial distribution of gas pressure, that is, steady volume currents. 
The divergence of this volume currents brings about a spatial distribution of field-
aligned currents, i.e. magnetospheric sources of ionospheric current systems. Such 
approach offers, among other things, a ''totally gratuitous'' explanation (and adequate 
description!) of the substorm ''breakup'' [Ponomarev, 1981]. We now consider this 
issue in slightly greater detail. It is known [Ponomarev, 1985] that the contents of the 
magnetic flux tube (MFT) to be referred to as the plasma tube (PT) throughout the text, 
transfers from one MFT to another in the convection process without surplus and 
deficiency in the case where the field lines of the magnetic flux tube are equipotential 
ones. This idealization is quite realistic everywhere apart from polar auroras.\\
     Then, as the PT drifting toward the Earth in a dipole field, its volume decreases in 
proportion to $L^{-4}$, and the situation is the reverse for density, while pressure increases 
in proportion to $\sim L^\frac {20}{3}$. However, the process of adiabatic compression is attended by 
the processes of PT depletion due to pitch-angle diffusion into the loss cone. This 
process is described by the factor
$\sim {exp}(- \int\ \frac{dt}{\tau})={exp}(-\int\ \frac{dr}{v_r\tau})={exp}(-\int\ \frac{rdv}{v_\vartheta \tau})$. 
Thus gas pressure has a maximum on each line of convection. In accordance with the 
equation for $p_g$ [Ponomarev, 1985], we have:
$$
P_g=P_g^0\left(\frac{L_\infty }{L}\right)^\frac{20}{3}{exp}\left(-\frac{5}{3}\int\ \frac{dr}{v_r\tau }\right) \eqno (1)
$$        
Here $p_g$ is gas pressure, L is the L-coordinate, $r = LR_e$ is the distance to the Earth ($R_e$
being the Earth's radius), $V_r$ and $V_\vartheta $ are the radial and azimuthal components of the 
convection velocity of the equatorial trace of the plasma tube, respectively, and $\tau$ is the 
characteristic time of PT depletion due to pitch-angle diffusion. The initial pressure at a 
certain boundary $L_\infty $ was considered time-independent in [Kennel, 1969]. For reasons 
unknown, Kennel did not extended his model to the unsteady-state case. This was done 
by one of us in [Ponomarev, 1981; Ponomarev, 1985; Anistratenko and Ponomarev, 
1981]. A typical gas pressure pattern that results through the combined action of 
convection and loses, is depicted in Fig. 1a. It has the form of an amphitheater with a 
clearly pronounced maximum near the midnight meridian, and with a sharp earthward 
"break". This ''break'' received the name ''Inner Edge of the Plasma Sheet'', IEPS.\\
     The projection of the ''amphitheater'' onto the ground corresponds to the form and 
position of the auroral oval. This projection, like the real oval, executes a motion with 
a change of the convection electric field, and expands with an enhancement of the field. 
In this process the amplitude at a maximum increases as the IEPS approaches the Earth. 
Next we consider the case where the boundary conditions in (1) are time-dependent. 
Let the pressure on the boundary be increased by, say, a factor of two. This ''impulse'' 
will start to drift downstream with the convection velocity, with a region of double 
amplitude remaining everywhere in its wake. If the ''impulse'' is of short duration, then 
a region ''multiplied by two'' of a limited size will travel downstream.  The effect of 
multiplication of two spatially narrow signals is always small apart from the time when 
their maximal coincide. An amplitude ''flare'' will occur then. Just this is the 
explanation for the ''substorm breakup'', a simple, logical corollary of the 
inhomogeneity of the system and motion [Ponomarev, 2000].\\
     Fig. 1b illustrates the second phase of development of the pressure pattern in the 
process of a model substorm.\\
     Based on the spatial distribution of pressure as a function of coordinates and time, 
we can calculate the spatial distribution of volume currents:
$$                                        
j=\frac{c[{\bf B}\times {\bf \nabla P_g]}}{B^2} \eqno (2)
$$

The divergence (2) under steady-state conditions gives an expression for field-aligned 
current densities:
$$
j_{\| }=cB_I \int_0 ^l \ \frac{[{\bf \nabla p_g \times \nabla p_B ] B}}{p_B B^3}\,dl \eqno (3)
$$ 
We perform the integration along a magnetic field line of the Earth's dipole field from 
the equator (0) to the ionosphere (l).
Noteworthy is the following property of the expression under the integral sign. It 
depends on the angle of intersection of magnetic and gas pressure contours. Within the 
dipole approximation $p_B ={const}$ are merely circles. On the contrary, $p_g ={const}$ have a 
complex configuration. The sign of current $j_{\| }$ depends, ultimately, on the sine sign of the 
angle between the normals to pressure contours. This factor eases qualitatively analysis 
of the current situation.

\section{ STATEMENT OF THE PROBLEM.}
Above we have outlined the prerequisites for the solution of the magnetosphere-
ionosphere coupling (MIC) problem in the part of it concerning their relations as the 
source and consumer of electric current and electric energy.\\
     The complexity of the MIC problem implies that currents in the ionosphere are 
governed by the electric field (with conductivity specified as a parameter), and in the 
magnetosphere they are determined by gas pressure gradient. There does exist a 
connection between the pressure distribution and convection, albeit relatively 
complicated. Our intention is to understand (by analyzing a maximum possible simple 
model that at the same time retains the most important traits of reality) how consistently 
current is established in the overall ionosphere-magnetosphere chain, how the 
magnetospheric generator of ionospheric currents operates, and what sources of power 
(including those of no electromagnetic origin) this generator uses to be at work. A 
partial answer to the last question has been given to date. We have demonstrated 
[Ponomarev, 1981; Ponomarev, 1985] that magnetospheric regions that operate like an 
MHD compressor where plasma is compressed under the action of Ampere's force 
$\frac{[{\bf j\times B}]}{c} $, satisfy the condition ${\bf V\nabla p_g}>0$, and regions where gas dynamic forces acts on 
electromagnetic forces, i.e. regions of MHD generators, satisfy the condition ${\bf V\nabla p_g}<0$ 
Conversion of energy from one kind to another may be written by a straightforward 
formula:
$$
{\bf V\nabla p_g }={\bf jE} \eqno (4)
$$                                  
It seems appropriate to employ in the analysis the region of the ''cleft'' which is 
produced when a plasma disturbance flows against the undisturbed pressure pattern (as 
a result of the unsteady-state character of boundary conditions as mentioned above). 
This detail of the pattern is clearly seen in Fig. 1b. Fig. 2 shows a schematic 
representation of a section of this pattern. The section of the cleft is represented by 
''corridors''. One can see that the walls of ''corridors'' serve as the sources of two 
bands of field-aligned currents which direction is opposite on different walls. On the 
whole, a current configuration forms, which corresponds to the Iijima-Potemra scheme 
[Iijima and Potemra, 1976]. Importantly, the stream convection lines run virtually along 
the axis of the "corridor", the "corridor" itself is extended with respect to the ... 
contours at a small angle, and hence the magnetic field inside it is nearly homogeneous. 
For that reason, the precipitation parameter $\tau $ can be considered a constant quantity.\\
     In the model of our interest, we replace the ''corridor'' itself by a rectangular 
channel with perfectly conducting walls overlaid by a conducting ''cover'', the 
ionosphere. We compensate for the difference in spatial scales, which is caused by the 
convergence of field lines, by a correction of parameters. The channel with a 
homogeneous magnetic field includes a steady flow of ideal plasma with a 
corresponding pressure gradient. All this is portrayed in detail in Fig. 3.

\section{MODEL OF THE SECONDARY MAGNETOSPHERIC GENERATOR.}

     Let us consider the phenomena occurring in the plasma ''corridor'' on the basis of a 
simple model. As is evident from Fig. 2, the orientation of the ''corridor'' is such that 
plasma flows nearly along its axis. The corridor is extended in a longitudinal direction; 
therefore, the magnetic field changes little within it. All these factors allow us to 
replace the ''corridor'' by a channel (extended along the axis Y) of width 2D, length L, 
and height H. The axis Y will be oriented across the channel, and the axis Z along its 
height, as shown in Fig. 3. The channel is filled with ideal plasma with pressure $p^0$ at 
the inlet and $p^1$ at the outlet. The magnetic field $B=(0,0,B_z)$ will be considered 
homogeneous. Plasma with the velocity $V=V_x (x)$ flows along the axis X in a positive 
direction. The walls of the channel possess infinite conductivity. The ionosphere is 
modeled by the upper cover of thickness h with Pedersen conductivity $\sigma$. As a 
consequence of the existence of a pressure gradient along the channel, the following 
current flows across it:
$$
j_y=\frac{c}{B} \left(\frac{\partial p}{\partial x}\right)
$$
To this volume density of current there corresponds the surface density and a total 
current:
$$
I_G (x)=\int j_y\ dz=Hj_y ,{ } J_G = \int I_G\  dx
$$
Accordingly, a total current of ionospheric load is:
$$
J_\sigma = \int\ \int \sigma E_1\  dx\prime  dz\prime = \sigma h \int E\  dx =\frac{\sigma hB}{c} \int V\ dx \eqno (5)
$$
The primes on the differentials signify that the integration is performed over the space 
of the ionosphere. Furthermore, because of the equipotentiality of magnetic field lines, 
the electric field in the ionosphere $E_I$ is related to the electric field in the 
magnetosphere by the relation: $E_Idx\prime = Edx$ In these formulas, c is the velocity of light.\\
     In addition to the current that closes through the ionosphere, a part of the MHD 
generator's current can close through the magnetosphere, as is the case with the 
corridor's current in Fig. 2. We designate this current by index 1. Then:
$$
J_1=\int\ \int j_{y1}\ dxdz=\int I_1\ dx
$$
From the condition of continuity of currents we find:
$$
\frac{dp}{dx}=-\frac{\sigma ^* B^2 V}{c^2}+\frac{I_1 B}{cH} \eqno (6)
$$
where $\sigma ^* =\sigma \left(\frac{h}{H}\right)$  .
The balance equation of gas kinetic energy in a steady-state one-dimensional case has 
the form:
$$
V\frac{dp}{dx}+\gamma p\frac{dV}{dx}=-\gamma \frac{p}{\tau} \eqno (7)
$$
Whence:
$$
p=p^0 \left(\frac{V_0}{V}\right)^\gamma {exp} \left(-\gamma \int\ \frac{dx}{V\tau}\right) \eqno (8)
$$
We now designate the initial level of gas pressure that is necessary and sufficient for 
supplying the ionosphere with electric current, by $p^{01}$ so that $p^0 = p^{01} + p^{02} $, where $p^{02}$ is 
the initial level of gas pressure that produces a current $J_1$.
$$
\gamma p^{01} \left(\frac{V_0}{V}\right)^{\gamma +1}{exp}\left(-\gamma \int\ \frac{dx}{V \tau}\right)\left[\frac{dV}{dx}+\frac{1}{\tau}\right]=\sigma ^* \left(\frac{B}{c}\right)^2 VV_0 \eqno (9)
$$
     
$$
-\gamma \left[ p^0 - p^{01} \left(\frac{V_0}{V} \right)^{\gamma +1} {exp} \left(-\frac{\gamma}{\tau}\int\ \frac{dx}{V}\right) \right] \left[\frac{dV}{dx}+\frac{1}{\tau}\right]=I_1 \left(\frac{B}{cH}\right)V_0 \eqno (10)         
$$

The solution of this system of equations that satisfies the conditions of our problem, is:
$$
V=V_0 - \frac{\gamma x}{(\gamma +2)\tau } \eqno (11)
$$
From (9) we obtain the condition:
$$
p^{01}=\frac{(\gamma +2)}{2\gamma}\left(\frac{B}{c}\right)^2 \sigma ^* \tau V_0^2 \eqno (12)
$$
And from (10) we get:
$$
I_1 =-\frac{2 \gamma}{(\gamma +2)}\left[p^0 -\frac{(\gamma +2)}{2\gamma }\left(\frac{B}{c}\right)^2\sigma ^* \tau V_0^2 \right]\frac{V}{B\tau V_0^2} \eqno (13)
$$
It is evident from (13) that the current $I_1$ is ''organized'' by the ''principle of balance'': 
all the necessary expenses of the ionosphere in current (power) are covered first, and 
what remains leaves for the geomagnetic tail region. As is evident from the figures, the 
current $I_1$ $(J_1)$ there becomes part of the dawn-dusk current. Only a part because there 
exists also the dawn-dusk current $J_B$ of a different origin. It is an external current with 
respect to the magnetosphere itself. As was shown by Ponomarev et al. [2000], it is 
produce at the Bow Shock (BS) front through a partial deceleration of solar wind 
plasma by Ampere's force with the involvement of this current. If the $B_z$-component of 
the Interplanetary Magnetic Field (IMF) is less than zero, the direction of this current is 
such that, by closing through the magnetospheric body, it produces there Ampere's 
force capable of acting to pushing magnetospheric plasma earthward, toward an 
increase of magnetic and gas pressure. Thus the MHD compressor lie in this region 
(located mostly at $5<L<10$ on the nightside, i.e. before the gas pressure maximum, see 
the figures; for details see in a book by Ponomarev [1985]. It is the gas compressed by 
the generator that is supplied to the MHD channel, the operation of which we are 
discussing here. Unlike the channel's region, the region of the MHD compressor lies in 
the area where the plasma is driven by magnetospheric convection to travel nearly 
radially to the Earth. From the balance of the gas pressure force and Ampere's force we 
have:
$$
J_1 +J_B = cH \int B^{-1}\ \left(\frac{dp}{dL}\right) dL \eqno (14)
$$
where $B=\frac{B_0}{L^3}$.
Whence:
$$
p^0 = q\left(\frac{B_c}{cH}\right)[J_1 +J_B]{,}  {where} {:} q=\frac{(4\gamma -1)}{4\gamma}\left(\frac{L_c}{L_T}\right)^{4\gamma}L_c^2 \eqno (15)
$$
$L_c$ and $L_T$ are the coordinates of the end and beginning of the area of plasma 
compression, and $B_c$ is the magnetic field strength at the compressor output. Further it 
will be assumed that $B_c = B$, that is, the MHD compressor output territorially coincides 
with the MHD generator input.
     Since plasma requires some time to travel the distance from the compressor input to 
output:
$$
\Delta T =\int_{L_T}^{L_c}\frac{R_e}{V_R}\ dL 
$$
then pressure at the MHD generator input will correspond to the earlier value of the 
compressor current.\\
     By integrating (13) over the entire length of the channel and assuming that the plasma 
velocity at the output is much smaller than that at the input of the MHD generator, we 
find:
$$
J_1=\left(\frac{cH}{B}\right)\left[p^0 -\frac{(\gamma +2)}{2\gamma}\left(\frac{B}{c}\right)^2 \sigma ^* \tau V_0^2 \right] \eqno (16)
$$
Upon substituting (15) into (16), in view of what has been said about the delay, we 
obtain an important relation:
$$
J_1 (t) -qJ_1 (t-\Delta T)=qJ_B (t- \Delta T) -J_{\sigma} (t) \eqno (17)
$$
In a steady state where there is no explicit time-dependence and $q = 1$:
$$
J_B = J_{\sigma} \eqno (18)
$$
This means that actually dissipative processes can take place in the magnetosphere only 
at the expense of an external source of current (and energy).
The whole of the complicated magnetospheric ''design'' only redistributes currents and 
energy fluxes in space and time.
Overall, though, this is an obvious inference as it is expectable. The integrity of (18) in 
this case implies that this is not merely a declaration now. We can point out the limits 
of applicability of (18) as well as the particular processes behind the notions ''steady 
state'' and ''unsteady state''.\\
     We now turn our attention to the ''cross-tail currents''. Let $J_1 +J_B$ be designated by $I_s$. 
Then from (17) it follows that:
$$
J_s (t)=qJ_s (t-\Delta T)+[J_B (t) -J_{\sigma} (t)] \eqno (19)
$$
Obviously, the control of the tail current $I_s$ proceeds both at the expense of a variation 
of $J_B$ and at the expense of the variation of the current of ionospheric load $J_\sigma$. In a quasi-
steady situation where $J^{-1}\frac{dJ}{dt}<<1 {,} q=1$ we have:
$$
\frac{dJ_s}{dt}\sim \frac{[J_B - J_{\sigma}]}{\Delta T} \eqno (20)
$$
Obviously, when $J_B>J_{\sigma}$, the cross-tail current increases, and the magnetospheric 
magnetic field is observed to extend into the tail. Otherwise when the ionospheric load 
current exceeds the external current,$dJ_s /dt <0$ and the tail current decreases, a 
''dipolization'' of the magnetic field occurs. The physical reason behind this is the 
increase in ionospheric consumption of current because of the increase in of 
conductivity caused by an enhancement of auroral particle precipitation.\\
     Thus between the consumer of current and energy, on the one hand, and their 
''general supplier'', the external current, there exists a flexible connection via a ''depot''
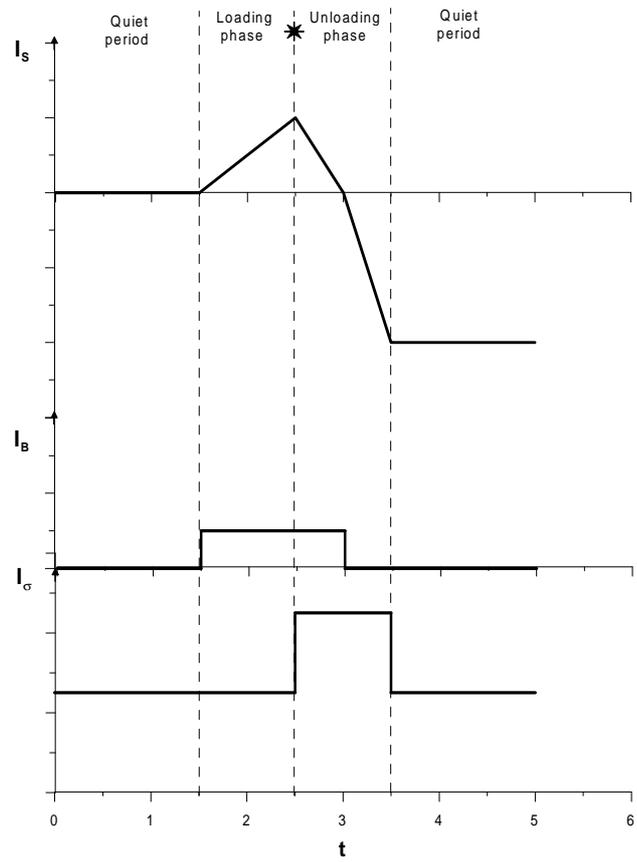
represented by current $J_1$. Fig. 4 presents the scheme of time response of currents to a 
change in integral ionospheric conductivity.

\section{CONCLUSION}
     We have shown that the consistent application of the idea put forward by Kennel 
[Kennel, 1969] which we further developed in [Ponomarev, 1981; Ponomarev et al., 
2000], makes it possible to successfully solve the magnetosphere-ionosphere coupling 
problem as regards the formation of auroral electrojets by volume currents generated in 
the magnetosphere by a corresponding distribution of plasma pressure. It was 
demonstrated that magnetosphere-ionosphere coupling mechanisms, along with the 
mechanisms of interaction between the magnetospheric MHD compressor and the MHD 
generator, only act to effect the redistribution of energy and electric current which must 
be supplied by external sources to feed the dissipative processes in the ionosphere 
Ponomarev et al. [2000] suggested a generation mechanism for this external current at 
the expense of a deceleration of solar wind plasma on the bow shock.\\
     We have been able, for the first time, to solve the problem of conjugacy of 
magnetospheric ''gradient'' current (dependent on plasma pressure gradient but 
independent on the electric field) with ''resistive'' ionospheric current dependent on the 
electric field (but independent of gas pressure). We pioneered the analysis of the 
combined operation of the magnetospheric MHD compressor and MHD generator 
which, in essence, represent a materialization of Kennel's idea of simultaneous 
existence of convection and precipitation in magnetospheric plasma.

\section{REFERENCES.}

- Anistratenko A.A. and Ponomarev E.A. Modeling of the formation conditions for 
particle precipitation areas and electric fields in the nightside polar ionosphere. 
Issledovaniya po geomagnetizmu, aeronomii I fizike Solntsa. Moscow: Nauka, v.53, 
pp.15-26, 1981.\\
- Axford W.I.  and  Hines C.O. A unified theory of high - latitude geophysical 
phenomena and geomagnetic storms. // Can. J. Phys., v. 39, p.1433, 1961. \\
- Dungey J.W. Interplanetary Magnetic Field and Auroral Zones. // Phys. Rev. Lett., 
v.6, p.47-48, 1961.\\ 
- Iijima T., Potemra T.A. The Amplitude Distribution of Field-Aligned Currents at 
Northern High Latitudes Observed by TRIAD. // J. Geophys.     Res., v.81, p. 2165-
2174, 1976.\\
- Kamide Y.  Substorm - 4. Results from ICS -4 // International SCOSTEP 
Newsletters, v.4, No.4 p.4, 1998.\\
- Kennel C.F. Consequences of magnetospheric plasma. // Rev. Geophys., 7, p. 379-
419, 1969.\\
- Lui A.T.Y.  Highlights on how to interpret auroral observations in terms of plasma 
sheet processes. // Proc. 5-th International Conference on Substorms, ESA SP-443, 
p.231, 2000.\\
- Mishin V.V. Dynamics and stability of shear flows on the boundaries of the 
magnetosphere, plasmasphere and solar wind. Author's Abstract of a Doctoral 
Thesis. Irkutsk, 2001.\\
- Ponomarev E.A. A model of the nightside ionosphere. Issledovaniya po 
geomagnetizmu, aeronomii i fizike Solntsa. Moscow: Nauka, v. 532, pp.3-14, 1981.\\
- Ponomarev E.A. Energy relations in the magnetosphere. Issledovaniya po 
geomagnetizmu, aeronomii i fizike Solntsa. Moscow: Nauka, v.53, pp.27-38, 1981.\\
- Ponomarev E.A. The Mechanisms of Magnetospheric Substorms. Moscow: Nauka, 
1985.\\
- Ponomarev E.A. On one plausible simple explanation for substorm break-up // 
Proc. 5 - th International conference on Substorms., ESA SP-443, à. 549, 2000.\\
- Ponomarev E.A., Urbanovich V.D., Nemtsova E.I. On the excitation  mechanism of 
magnetospheric convection by the Solar Wind. // Proc. 5 - th International 
conference on Substorms., ESA SP-443, à. 553, 2000.\\

\end{document}